\documentclass[preprint,showpacs,preprintnumbers,amsmath,amssymb]{revtex4}
\newcommand{\W}{14cm}

\usepackage{graphicx}
\usepackage{dcolumn}
\usepackage{bm}

\begin{document}

\preprint{}
\title{Fractal fronts in fractal fractures:  large and small-scale structure}

\author{G. Drazer}
\affiliation{Benjamin Levich Institute and Department of Physics,
City College of the City University of New York, New York, NY 10031}
\author{J. Koplik}
\affiliation{Benjamin Levich Institute and Department of Physics,
City College of the City University of New York, New York, NY 10031}
\author{H. Auradou}
\affiliation{Laboratoire Fluide, Automatique et Syst{\`e}mes Thermiques,
UMR No. 7608, CNRS, Universit{\'e} Paris 6 and 11,
B{\^a}timent 502, Universit{\'e} Paris Sud, 91405 Orsay Cedex, France.}
\author{J.P. Hulin}
\affiliation{Laboratoire Fluide, Automatique et Syst{\`e}mes Thermiques,
UMR No. 7608, CNRS, Universit{\'e} Paris 6 and 11,
B{\^a}timent 502, Universit{\'e} Paris Sud, 91405 Orsay Cedex,
France.}
\date{\today}

\begin{abstract}
The evolution and spatial structure of displacement fronts in fractures with self-affine 
rough walls are studied by numerical simulations. The fractures are open and the two faces are
identical but shifted along their mean plane, either parallel or perpendicular to the flow.
An initially flat front advected by the flow is progressively distorted into a self-affine 
front with Hurst exponent equal to that of the fracture walls.  The lower cutoff of the
self-affine regime depends on the aperture and lateral shift, while the upper cutoff grows 
linearly with the width of the front.
\end{abstract}
\maketitle

Fractures are ubiquitous in the Earth's upper crust and in many 
materials of engineering interest, and their size
varies over a broad range of length scales, from several meters in the
case of faults down to the sub-micron scale of micro-cracks.  
The geometry of the fracture obviously influences any transport process within
it, and in particular the hydraulic properties of rock fractures, 
crucial in such applications 
as hydrocarbon recovery, subsurface hydrology, and earthquake prediction.
In these situations, transport generally involves flow through entire
networks of fractures, and often through a porous matrix as well, 
but frequently 
the flow is dominated by {\em channeling} through  particular fractures
\cite{msb84,hl96}, and therefore there is considerable interest in
{\em fully} understanding flow through an individual fracture.\\
A key feature characterizing the pore space geometry
of fractured materials is the experimental observation that the roughness
of fracture surfaces is statistically described as a self-affine fractal structure
over an appreciable range of length scales
\cite{book-feder88,book-mandelbrot82}. This observation  holds 
for both man-made \cite{mpp84,blp90,khw93} and natural fractures
\cite{brown85,psj92}, and for a broad variety of materials and length
scales. We consider rock fracture surfaces without overhangs, whose height
can be described through a single valued function $z(x,y)$, where $(x,y)$
are Cartesian coordinates in 
the mean plane of the surface.  Self-affinity means
that the surface remains statistically invariant under the scale
transformation $z(\lambda x,\lambda y)=\lambda^\zeta z(x,y)$, where
the Hurst exponent $\zeta$ characterizes the roughness of the surface, in that
the fluctuation of the surface heights over a length $L$ is given by
$\sigma_z(L)=\ell \left(L/\ell \right)^\zeta$. Here $\ell$ is the topothesy, 
defined as the horizontal distance over which
fluctuations in height have a rms slope of one \cite{psj92}.  Experimental
values of $\zeta$ are found to be close to $0.8$, for most of the materials
considered and independent of the fracture mode
\cite{mpp84,blp90,khw93,brown85,psj92}, with the exception of materials displaying
intergranular fractures, such as some sandstone rocks, for which it has
been found to be close to $0.5$ \cite{bah98}.
The same exponent $\zeta$ describes the scaling behavior of the two-point
height correlation function, and therefore, 
one expects the fractal geometry of the surfaces to induce spatially correlated 
velocity fluctuations.\\ 
A particularly sensitive measurement of the
correlated character of the velocity fluctuations is the shape of an advected 
front of tracer particles, because 
each particle position is an integral over the velocity field along its path. 
Recent displacement experiments have found a striking approximate correlation 
between tracer motion and surface geometry
--  the shape of the front was found to be self-affine, with an exponent 
close to the Hurst exponent $\zeta = 0.8$ of the fracture wall \cite{ahr01}. 
The objective of this paper is to establish this connection in quantitative 
detail, and determine its range of validity and further implications, 
using Lattice-Boltzmann (LB) numerical simulations 
\cite{long_paper}.  
We consider the motion of passive tracers in single fractures, which are modeled 
as the opening between two perfectly matching self-affine rough surfaces.
We assume that there is no contact between the two 
fracture surfaces, and that the walls are chemically inert, completely rigid, and non-porous.
As in the experiments motivating this work, we focus on the case of small Reynolds
numbers for which inertia effects can be neglected, and P\'eclet number sufficiently high
to ignore molecular diffusion.\\  
\begin{figure}
\includegraphics[width=\W]{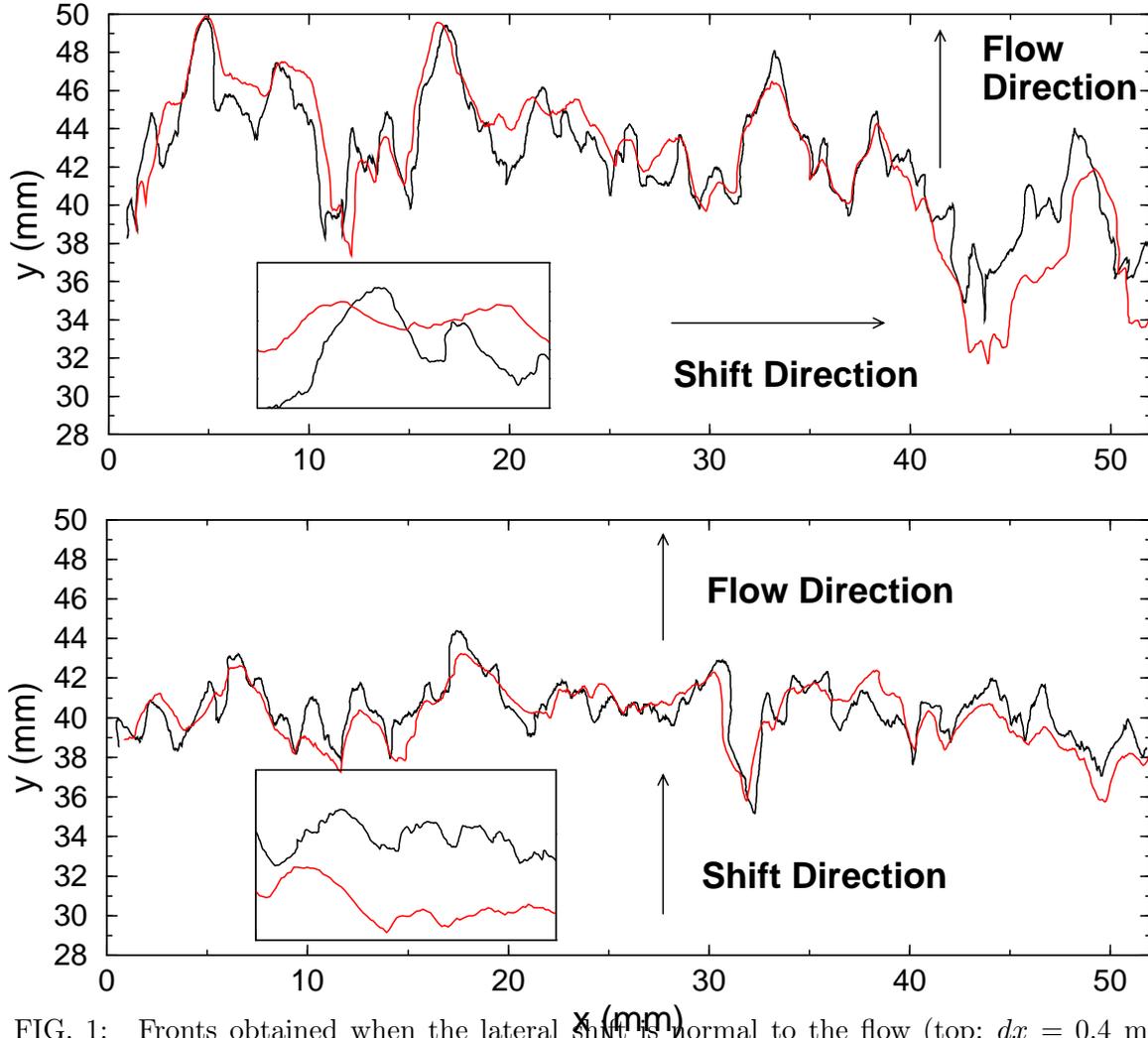}
\caption{\label{fig:figure1} Fronts obtained when the lateral shift
is normal to the flow (top; $dx=0.4\ \rm{mm}$) and parallel to the flow (bottom; $dy=0.4\ \rm{mm}$).  
The black lines correspond to a fracture with Hurst exponent $\zeta=0.5$ and the gray lines
correspond to $\zeta=0.8$. The inserts display $4\ \rm{mm} \times 4\ \rm{mm}$ sections of the fronts.} 
\end{figure}
A self-affine surface is first generated numerically, using a Fourier
synthesis method \cite{dk02}, and the fracture pore space
is the region between one such surface and a suitably shifted replica.
First, the two surfaces are
separated by a fixed distance $h$, in the direction normal to the mean
plane of the fracture, and then a lateral shift of the upper surface
is introduced.  The effects of the latter lateral shift, meant to reflect subsurface 
motion during or subsequent to fracture, are an important practical issue.
The maximum allowed shift corresponds to the occurrence
of the first contact point between the two surfaces, so that the fracture is open
over its full area. Two orthogonal shift directions along the
mean plane of the fracture are considered; $dy$, in the direction of the mean flow, 
and $dx$, perpendicular to it.  
The free space between the two surfaces is
occupied by a cubic lattice of fixed lattice spacing, 
with $1024 \times 1024 \times 20$ sites, which is 
large enough to accurately capture the three-dimensionality 
of the velocity field \cite{dk02}.  

In order to generate numerically fractures that are comparable to
those used in the experiments, we need to match the
two relevant length scales, the topothesy and the mean aperture,
as well as the Hurst exponent. Thus, the mean aperture is set to $1\
\rm{mm}$, corresponding to a fracture size $L=51.2\ \rm{mm}$ ($\delta = 0.05\ \rm{mm}$), 
equal to one fourth of the fracture used  in \cite{ahr01}. 
In order to have the same surface roughness,
the amplitude of average fluctuations in surface height over a length of $1\ \rm{mm}$ is set
to $0.25\ \rm{mm}$.  Two different values of the Hurst exponent $\zeta$ are 
used here:  $0.8$ and $0.5$.  The first value corresponds to the measured
Hurst exponent of the material used in \cite{ahr01}, and the second one
is typical of fractured materials with intergranular effects.  
Considerable height variations occur over a distance comparable to the aperture,
$\sigma_z(h) = 0.25\ \rm{mm}$ and, therefore, the lubrication approximation, which assumes a 
Poiseuille velocity profile across the aperture,  fails to describe the flow field inside the fractures \cite{mta95}. 
Therefore, the three-dimensional flow field in the pore space of the fractures is calculated, for different values
of the lateral shift, using a LB method \cite{qhl92}.  
Periodic boundary conditions are used in the $x$ and $y$ directions, including the
inflow and outflow, and the Reynolds number is small ($\sim 0.2$).  

\begin{table}
\begin{tabular}{c|c|c||c|c|c}
 $\zeta=0.5$     & $dy=0.4$ & $dx=0.4$ & $\zeta=0.8$ & $dy=0.4$ & $dx=0.4$\\
 \cline{1-6}
 $\langle y \rangle$ & $40.40$ & $42.56$ & $\langle y \rangle$ & $40.1$ & $42.4$\\
 $\sigma$ & $1.43$ & $3.11$ & $\sigma$ & $1.55$ & $4.25$\\
 $e$ & $9.23$ & $15.82$ & $e$ & $7.49$ & $18.20$
 \end{tabular}
 \caption{\label{tab:tab1}Statistical properties of the front in $\rm{mm}$ displayed
 in Fig.~\ref{fig:figure1};  $\langle y \rangle$ is the mean
 position of the tracers in the flow direction, $\sigma=\langle(y-\langle y \rangle)^2 \rangle^{1/2}$,
 and $e=y_{max}-y_{min}$} 
\end{table} 

 \begin{figure}
\includegraphics[width=\W]{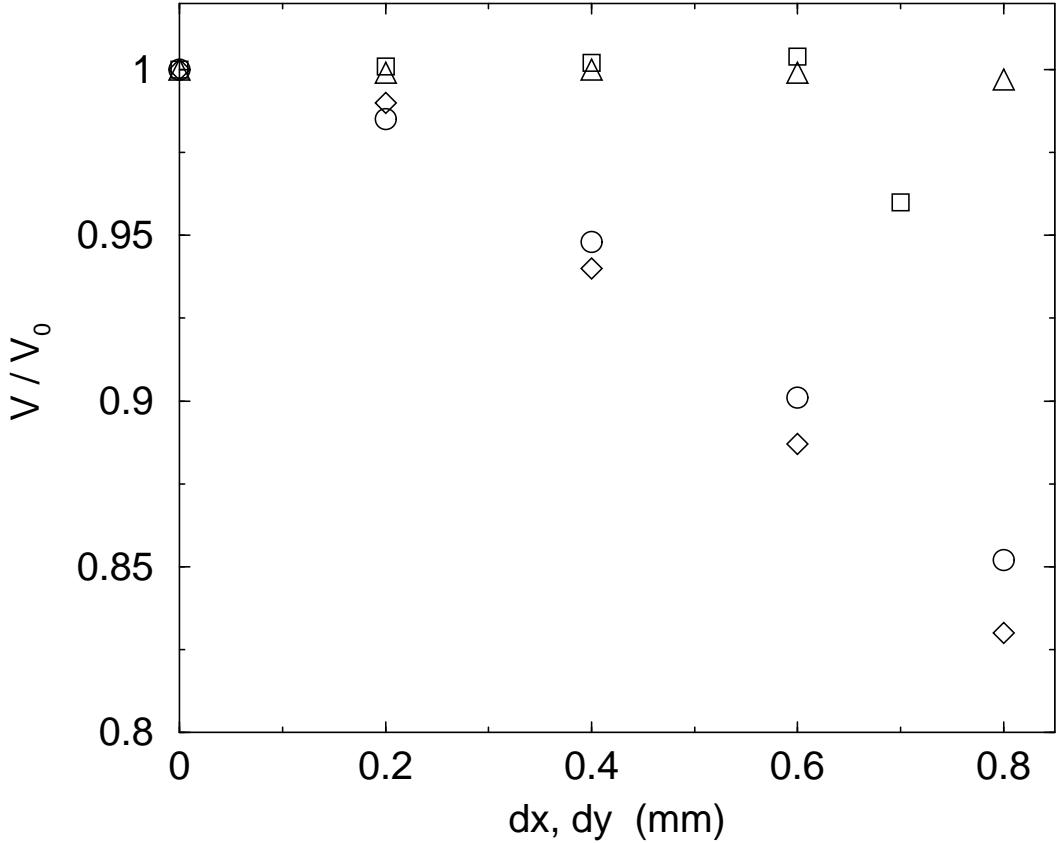}
\caption{\label{fig:figure2} Mean velocity of the tracers normalized by the
velocity for zero shift as function of the shift. $\circ$ and $\diamond$ are for flow along the shift direction within fractures of wall roughness $\zeta=0.5$ and $0.8$. $\square$ and $\triangle$ correspond to flow normal to the shift in fractures with $\zeta=0.5$ and $0.8$.}
 \end{figure}

In the displacement experiments presented in Ref. \cite{ahr01}, a steady
radial flow of a clear fluid is first established in a transparent 
fracture model, cast in epoxy from natural granite. 
The same fluid, with a small concentration of a dye, 
is then injected and the spreading front of the dye is
recorded. 
Several mechanisms contribute to the dispersion of tracer particles,
among them molecular diffusion and Taylor dispersion. However, 
the most relevant process that drives the spreading within fractures is the 
velocity fluctuations induced by the surface roughness \cite{ahr01}. 
This mechanism, known as
{\it{geometric}} or {\it macro}-dispersion, is linear in the velocity,
and therefore leads to an external front that is independent of the
flow rate for a fixed injected volume. The front is thus essentially
determined by the {\em two dimensional} geometrical disorder of the 
velocity field in
the mean fracture plane. Comparisons between experimental and numerical
dispersion results can thus be achieved by considering the spreading
of tracer particles moving in the flow field resulting from 
averaging the three dimensional velocities over the gap of the fracture.

Initially in the computations, a tracer particle is located at each
of the $1024$ lattice sites corresponding to the injection side of the
fracture. Each particle then propagates in the gap averaged velocity field
and the velocity fluctuations due to the effective aperture variation
will distort the initially flat front.  Figure \ref{fig:figure1} 
shows typical invasion fronts obtained after the tracer traverses roughly 
$80\%$ of the system.  
An obvious qualitative feature is that 
the front is rougher when the relative shift of the fracture surfaces 
is normal to the mean flow.  Quantitatively, these small-scale features 
vary significantly with both the Hurst exponent and the shift direction, 
and in particular, the front contains more small scale details when the 
fracture wall roughness is decreased.  
The basic numerical properties of these fronts are entirely consistent with 
their visual appearance, as shown in Table \ref{tab:tab1}.
We will show presently that
these fronts are in fact fractals, over a range of length scales. 
In addition, in Fig. \ref{fig:figure2}, we show that there is a remarkable anisotropy in 
the average propagation velocity of the front, in that a relative shift perpendicular 
to the mean flow has a negligible effect, whereas a shift along the flow produces a systematic
decrease in the permeability.
Qualitatively, this behavior may be understood 
from the observation that transverse shifts simply replace one preferred flow
channel with another of comparable hydraulic resistance
while longitudinal shifts tend to block the flow,
but the systematic variation, or lack of it, seen here is quite surprising \cite{dk02}. 

\begin{figure}
\includegraphics[width=\W]{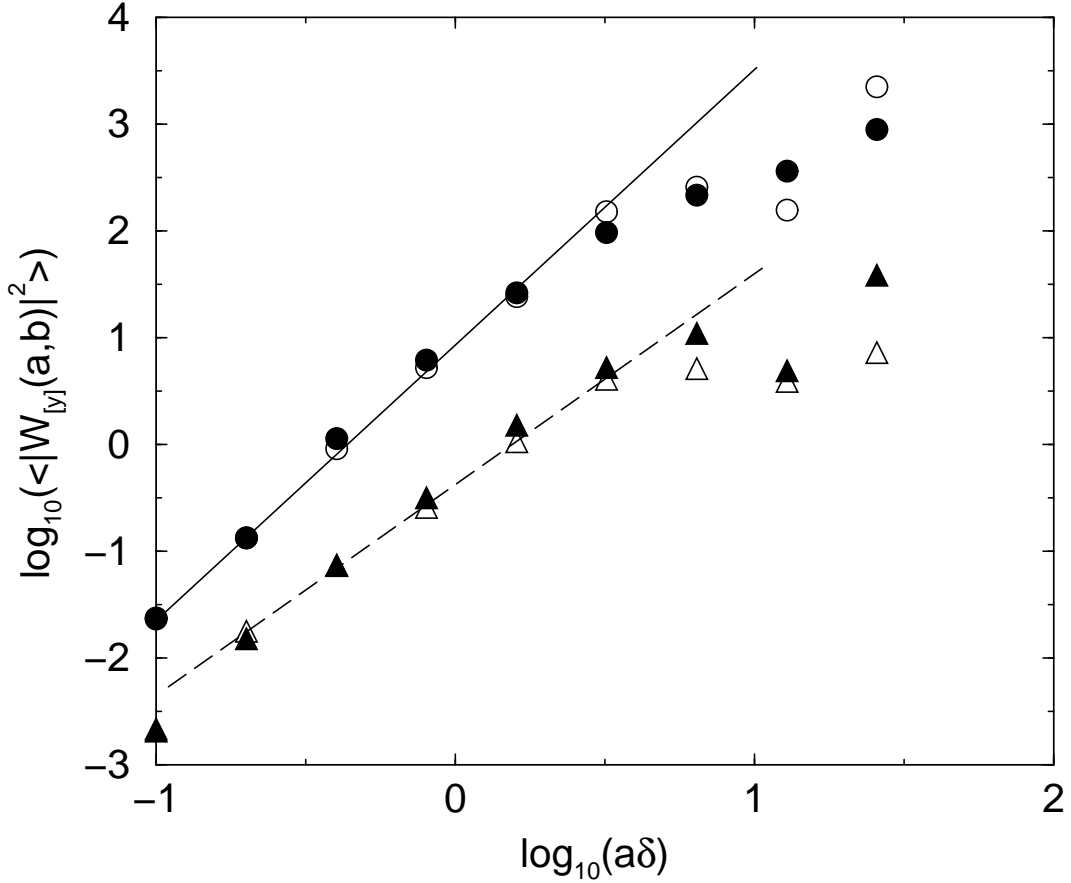}
\caption{\label{fig:figure3} Variation of
$log_{10}(\langle |W_{[y]}(a,b)|^2\rangle_b)$ as function of  $log_{10}(a\delta )$. 
Circles and triangles correspond to $\zeta=0.8$ and $\zeta=0.5$ respectivley.
Solid and open symbols correspond to $dy=0.4$ and $dx=0.4$ respectively.
The two data sets are shifted vertically for convenience. The
solid line has a slope of $2.6$ while the dashed line has  a slope of $2$.}   
\end{figure}

\begin{figure}
\includegraphics[width=\W]{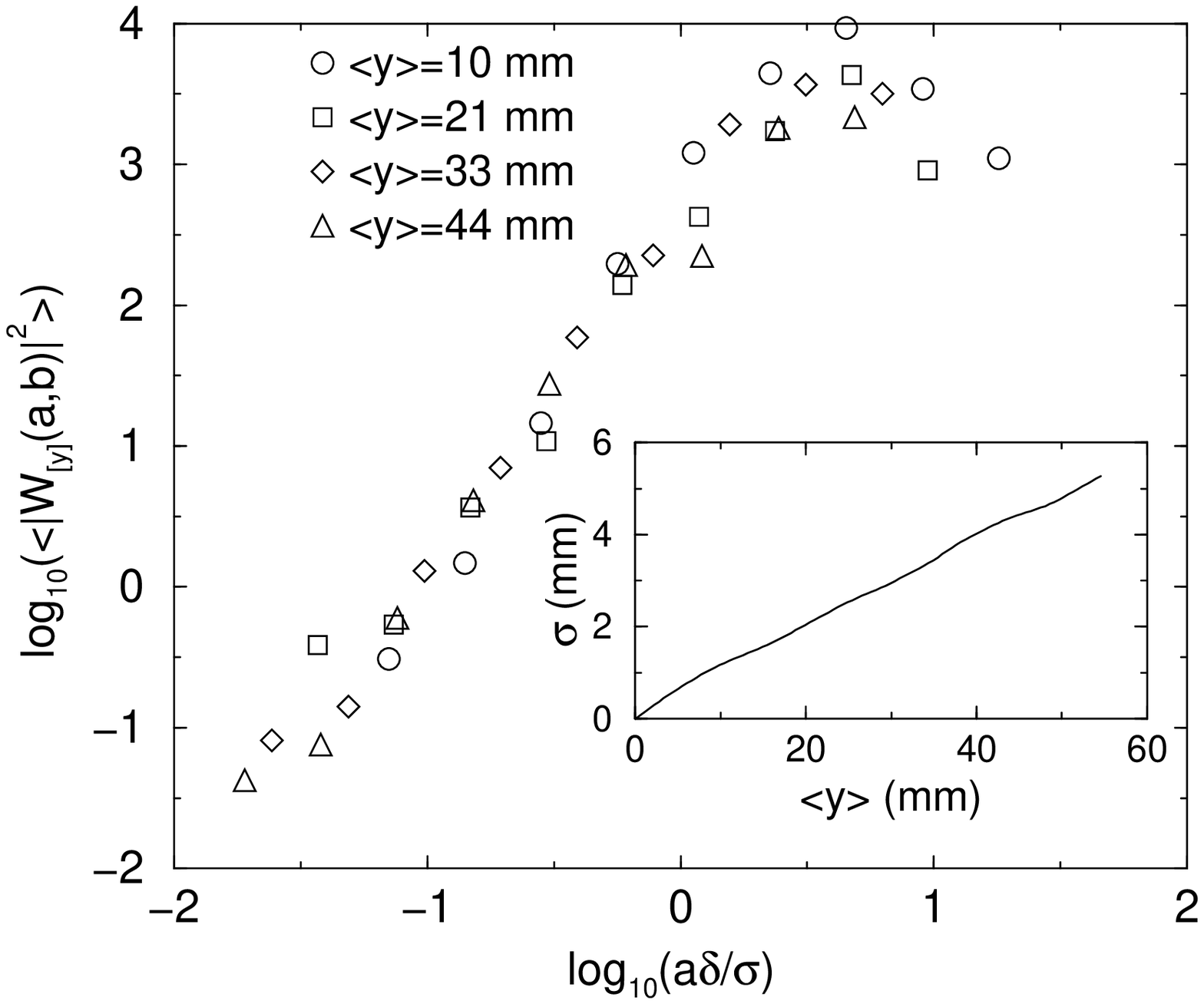}
\caption{\label{fig:figure4} $log_{10}(\langle |W_{[y]}(a,b)|^2\rangle_b)$ as
function of  the normalized scale parameter
$log_{10}(a\delta/\sigma)$. The data set are shifted vertically to make comparisons
easier. They
 correspond to fronts obtained  after different 
mean traveled distances $\langle y \rangle$, in a fracture of roughness $\zeta=0.8$ 
and with walls shifted by $dx=0.4\rm{mm}$. 
The insert shows the variation of $\sigma$ as a function of the mean traveled distance.}  
\end{figure}

The experimental results presented in \cite{ahr01} were analyzed for 
their fractal dimension $D$ using the average mass method \cite{book-feder88},
in which a constant density of tracer is assumed for the invasion
front and the average number of tracer particles, $\langle M(r) \rangle$,
located at a distance less than $r$ from a random origin on the contour 
is computed. For a fractal front, a power law relationship $\langle
M(r) \rangle \propto r^D$ is expected, where $D$ is the fractal dimension
ranging between $1$ and $2$, and if the front is furthermore assumed to be 
self-affine, one would identify $\zeta_f=2-D$.  The result found there was 
$\zeta_f = 0.74 \pm 0.03$, close to the Hurst exponent of the fractures, $ \zeta=0.8$.
If we apply the same analyzis to the numerical results here, we obtain \cite{long_paper} 
$\zeta_f=0.82\pm 0.01$ when $\zeta=0.8$ 
and $\zeta_f=0.5 \pm 0.15$ for $\zeta=0.5$,  for length scales between 
certain upper and lower cutoffs.
The lower cutoff is of the order the mean aperture, and independent
of the direction and magnitude of the lateral shift, 
as were the dimensions $\zeta_f$ themselves.  However, the upper
cutoff is shift-dependent, and difficult to interpret.
An additional and severe shortcoming of this procedure is 
that it does not directly test self-affinity. 

A more appropriate method for self-affine structures is the average  wavelet coefficient 
(AWC) analysis \cite{simonsen98}, where
a one dimensional front $y(x)$ is transformed into the wavelet domain as
\begin{equation}
W_{[y]}(a,b) = \frac{1}{\sqrt{a}} \int \psi^*_{a,b}(x) y(x) dx,
\end{equation}
where $\psi_{a,b}(x)$ is obtained from the {\it analyzing wavelet} $\psi$, in our case a Daubechies
wavelet \cite{Daubechies}, via rescaling and translation, $\psi_{a,b} (x) = \psi((x-b)/a)$.  
The AWC measures the average ``energy'' present in the front at a given scale, 
defined as the arithmetic average of $|W_{[y]}(a,b)|^2$ over all possible
locations $b$, and for a statistically self-aff{\rm }ine profile with 
exponent $\zeta_f$, it scales as $\langle |W_{[y]}(a,b)|^2\rangle_b \sim
a^{2\zeta_f + 1}$. Typical results of applying the AWC method to the 
numerical fronts shown in Fig.~\ref{fig:figure1} are given in 
Fig.~\ref{fig:figure3}, and we see a robust scaling behavior, 
depending only on the roughness  exponent of the fracture surfaces and independent of the
relative shift and the invasion time.  We find $\zeta_f=0.82\pm0.1$ for 
surfaces with roughness  $\zeta=0.8$ and $\zeta_f=0.5 \pm 0.15$ for surfaces with $\zeta=0.5$, 
consistent with the values found using the average-mass method.
These results are in good agreement with
previous experimental studies of tracer dispersion \cite{ahr01} and
with a theoretical analysis based on a perturbative approach \cite{rph98}.

\begin{figure}
\includegraphics[width=\W]{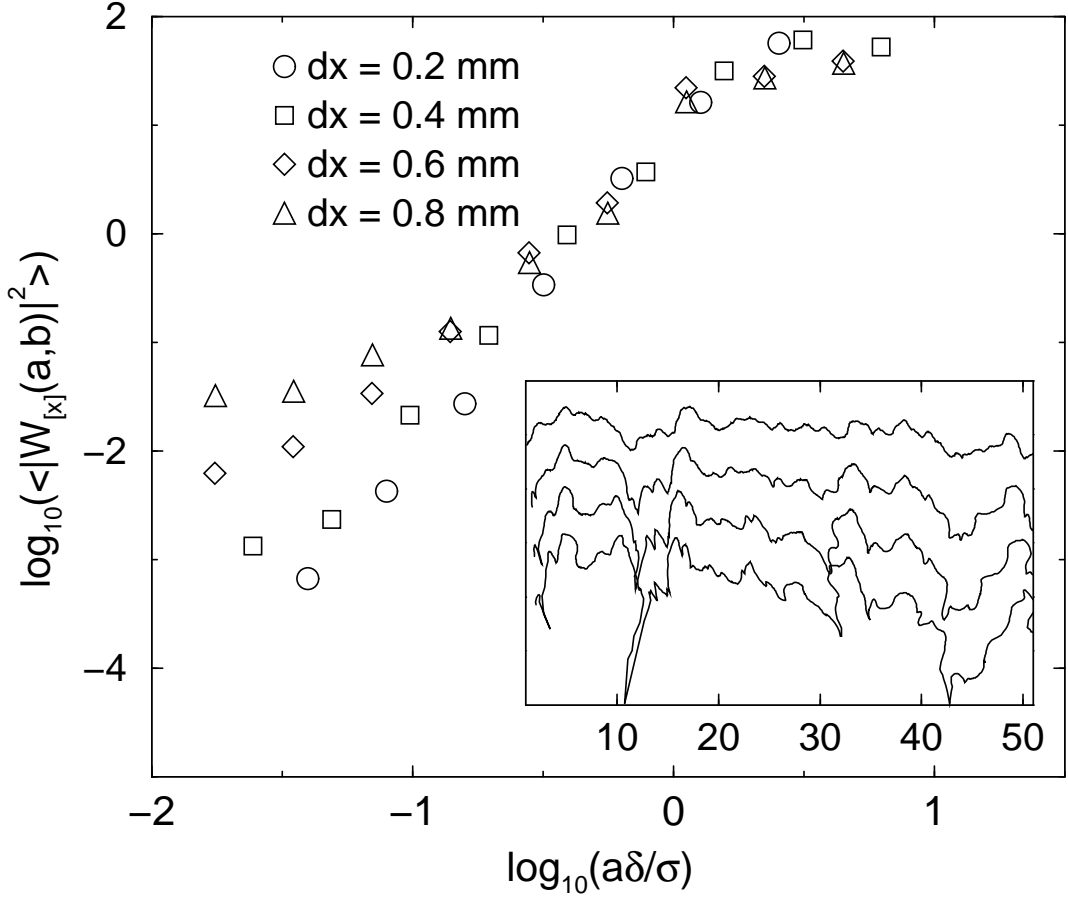}
\caption{\label{fig:figure5}  Variation of
$log_{10}(\langle |W_{[y]}(a,b)|^2\rangle_b)$ as function of  
$log_{10}(a\delta/\sigma)$. The data sets are shifted vertically to make comparisons
easier. They
correspond to fronts obtained for the same average traveled distance 
$\langle y \rangle = 40.7\ \rm{mm}$ and for the same fracture ($\zeta=0.8$), 
but for different values of the lateral shift. 
The insert displays the corresponding fronts for increasing shifts from top to bottom. } 
\end{figure}

As usual, one sees scaling behavior only between certain cutoff length 
scales. The origin of the upper cutoff can be understood by considering 
the ``energy'' as a function of $a\delta /\sigma$, 
where $\sigma$ is the width of the front.  
In Fig.~\ref{fig:figure4}, data for the case $\zeta=0.8$ and
$dx=0.4$ are plotted at different times, showing that the upper cutoff
for scaling behavior is proportional to the width of the front.
The width itself grows roughly linearly as a function of time, or 
equivalently with the mean distance traversed by the front, as seen in 
the inset of Fig.~\ref{fig:figure4}, so the upper limit of self-affine behavior
increases with time. To understand the lower cutoff, it is instructive to compare fronts at
the same mean traversed distance for different values of the shift.
In Fig.~\ref{fig:figure5} we display data for the $\zeta=0.8$ fracture
with different values of lateral shift, at the same transit time; 
the lower cutoff is seen to increase with increasing shift.
The connection between scaling breakdown and shift can be elucidated by
simply examining the front shapes in these four cases, as shown in
the inset to Fig.~\ref{fig:figure5}.  An increasing shift tends to produce  
stagnant low-velocity regions where the front halts, and these
regions destroy the self-similarity.  The AWC method is particularly
well-suited to address the lower cutoff issue, because the stagnant tails of
the front contribute to the wavelet ``energy'' strongly at low values of
$a$.  In the average-mass method, in contrast, 	they contribute
predominantly at large $r$ whereas the lower cutoff is 
on the order of the mean aperture, independent of the surface
exponent and any lateral shift \cite{ahr01,long_paper}.  

There is a an obvious intuitive appeal to the idea that a fractal geometry
will somehow impose itself on a suitably flexible object propagating 
through it. In this paper we have used numerical simulations to expand 
upon the experimental observation that a (high P\'eclet number) 
displacement front in a self-affine fracture behaves in this way.
We have presented compelling evidence that self-affine fronts indeed appear
in this situation, for a wide range of parameters, 
and have further determined the relevant range of length scales for
scaling behavior.  This work raises a number of further intriguing 
questions, including (1) The behavior of higher moments of the front geometry;
(2) The interplay between the self-affine character and the strong anisotropic
behavior in {\it radial} injection flows; (3) The effects of finite 
P\'eclet numbers, which would presumably alter the scaling cutoffs lengths, 
on hydrodynamic dispersion.
We hope to address these and other questions in future work.

We have benefited from discussions with S. Roux.
HA and JPH are supported by the CNRS and ANDRA through the GdR FORPRO (contribution No. $2003/12A$). 
This work was partly supported by the ECODEV and PNRH CNRS
 programs. 
GD and JK are supported by the Geosciences Program of the Office of Basic 
Energy Sciences, US Department of Energy. Computer resources were provided by the National Energy Research Scientific Computing Center. This work was facilitated by a 
CNRS-NSF Collaborative Research Grant.


\begin{thebibliography}{10}

\bibitem{msb84} 
C.A. Morrow and L.Q. Shi, J. Phys. A {\bf 84}, 3193 (1984).

\bibitem{hl96} 
E. Hakami and E. Larsson, Int. J. Rock Mech. Min. Sci. \& Geomech. Abstr.
{\bf 33}, 395 (1996).

\bibitem{book-feder88}
J. Feder, {\em Fractals} (Plenum, New York, 1988).

\bibitem{book-mandelbrot82}
B. Mandelbrot, {\em The Fractal Geometry of Nature} (W.H. Freeman, New York,
  1982).

\bibitem{mpp84}
B.B. Mandelbrot and D.E. Passoja and A.J. Paullay, Nature {\bf 308}, 721
(1984).

\bibitem{blp90}
E. Bouchaud, G. Lapasset, and J. Plan{\`e}s, Europhys. Lett. {\bf 13},  73  (1990).

\bibitem{khw93}
J. Kertsez, V.K. Horv\'ath and F. Weber, Fractals {\bf 1}, 67 (1996).

\bibitem{brown85}
S. Brown and C. Scholz, J. Geophys. Res. {\bf 90},  12575  (1985).

\bibitem{psj92}
C. Poon, R. Sayles, and T. Jones, J. Phys. D: Appl. Phys. {\bf 25},  1269
  (1992).

\bibitem{bah98}
J.M. Boffa, C. Allain and J.P. Hulin, European Phys. J. - App. Phys. {\bf
2}, 2 (1998).

\bibitem{mta95}
V.V. Mourzenko, J.-F Thovert and P.M. Adler, J. Phys. II France {\bf 5}, 465
(1995).

\bibitem{ahr01}
H. Auradou, J.P. Hulin and S. Roux, Phys. Rev. E {\bf 63}, 066306 (2001).

\bibitem{long_paper} We have used particular examples taken from extensive
numerical simulations to draw our conclusions;  further details and analysis
will be provided in a longer paper now in preparation.

\bibitem{dk00}
G. Drazer and J. Koplik, Phys. Rev. E {\bf 62}, 8076 (2000).

\bibitem{dk01}
G. Drazer and J. Koplik, Phys. Rev. E {\bf 63}, 056104 (2001).

\bibitem{dk02}
G. Drazer and J. Koplik, Phys. Rev. E {\bf 66}, 026303(2002).

\bibitem{qhl92}
Y.H. Qian and D. d'Humieres and P. Lallemand, Europhys. Lett. {\bf 17}, 479
(1992). 

\bibitem{simonsen98}
I. Simonsen, A. Hansen and O.M. Nes, Phys. Rev. E. {\bf 58}, 2779 (1998).

\bibitem{Daubechies}
I. Daubechies, {\it Ten Lectures on Wavelets} (SIAM, Philadelphia, 1992).

\bibitem{rph98}
S. Roux, F. Plourabou\`e and J.P. Hulin, Transp. Porous Media {\bf 32}, 97
(1998). 


\end{thebibliography}
\end{document}